# Large bandgap quantum anomalous hall insulator in a designer ferromagnet-topological insulator-ferromagnet heterostructure

Qile Li[1,2,3], Chi Xuan Trang[1,2], Weikang Wu[4,5], Jinwoong Hwang [6], Nikhil Medhekar[2,3], Sung-Kwan Mo[6], Shengyuan A. Yang[4], Mark T Edmonds[1,2*]

[1]School of Physics and Astronomy, Monash University, Clayton, VIC, Australia.

[2]ARC Centre for Future Low Energy Electronics Technologies, Monash University, Clayton, VIC, Australia.

[3]Department of Materials Science and Engineering, Monash University, Clayton, VIC, Australia

[4]Research Laboratory for Quantum Materials, Singapore University of Technology and Design, Singapore 487372, Singapore

[5] Division of Physics and Applied Physics, School of Physical and Mathematical Sciences, Nanyang Technological University, Singapore 637371, Singapore

[6]Advanced Light Source, Lawrence Berkeley National Laboratory, Berkeley, CA, USA.

* Corresponding author: mark.edmonds@monash.edu

**Abstract -** Combining magnetism and nontrivial band topology gives rise to quantum anomalous Hall (QAH) insulators and exotic quantum phases such as the QAH effect where current flows without dissipation along quantized edge states. Inducing magnetic order in topological insulators via proximity to a magnetic material offers a promising pathway towards achieving QAH effect at high temperature for lossless transport applications. One promising architecture involves a sandwich structure comprising two single layers of $MnBi_2Te_4$ (a 2D ferromagnetic insulator) with ultra-thin $Bi_2Te_3$ in the middle, and is predicted to yield a robust QAH insulator phase with a bandgap well above thermal energy at room temperature (25 meV). Here we demonstrate the growth of a 1SL $MnBi_2Te_4$ / 4QL $Bi_2Te_3$ /1SL $MnBi_2Te_4$ heterostructure via molecular beam epitaxy, and probe the electronic structure using angle resolved photoelectron spectroscopy. We observe strong hexagonally warped massive Dirac Fermions and a bandgap of $75 \pm 15$ meV. The magnetic origin of the gap is confirmed by the observation of broken time reversal symmetry and the exchange-Rashba effect, in excellent agreement with density functional theory calculations. These findings provide insights into magnetic proximity effects in topological insulators, that will move lossless transport in topological insulators towards higher temperature.

# Introduction

Three-dimensional (3D) topological insulators (TIs) are materials with non-trivial band topology that are distinct from regular band insulators. Due to strong spin-orbit coupling, band inversion and bulk boundary correspondence, the surfaces of 3D TIs host spin polarised massless Dirac cones that are robust against time reversal symmetric perturbations[1-3]. By introducing long-range magnetic order in the 3D TI, time reversal symmetry (TRS) can be broken, which manifests as a gap opening in the Dirac cones due to exchange interaction and gives rise to novel phases such as the quantum anomalous Hall (QAH) insulator and axion insulator[4-6]. A QAH insulator possesses chiral edge states within the band gap, and when the chemical potential is tuned into the magnetic gap a quantised Hall conductance and almost zero longitudinal resistance is observed known as the quantum anomalous Hall effect [7,8]. Characterised by dissipation-less charge transport and perfect spin polarisation, QAH insulators have potential applications in lossless transport, spintronics and topological quantum computing involving braiding of Majorana Fermions[9].

To date, the QAH effect has been observed in dilute magnetically doped ultra-thin TIs [7,8] at 30 mK, and up to 1 K in modulation-doped sandwich heterostructures[10], and most recently in five-layer exfoliated flakes of the intrinsic magnetic topological insulator, $MnBi_2Te_4$ at 1.4 K[11]. However, the temperature at which the QAH effect is observed in these systems is well below the size of the magnetic gap and the Curie temperature. This is a result of magnetic disorder[12] and the difficulty in incorporating 3$d$ transition metal magnetic dopants.

Rather than incorporating 3$d$ transition metals into the crystal lattice, a more advantageous strategy may be to place two ferromagnetic materials on the top and bottom surfaces of a 3D TI. This will break TRS for topological surface states (TSS) at the two surfaces *via* magnetic proximity, thereby opening an exchange gap and giving rise to a QAH insulator or axion insulator phase [13,14] (depending on the thickness of the TI layer). However, inducing sufficient magnetic order to open a sizeable exchange gap *via* proximity effects is challenging due to the undesired influence of the abrupt interface potential as a consequence of lattice mismatch between the magnetic material and TI[15]. To date, the only demonstration of QAH effect in a ferromagnetic insulator (FMI) – topological insulator (TI) – ferromagnetic insulator (FMI) heterostructure has been in $(Zn_{1-x}Cr_x)Te$-$(Bi_{1-y}Sb_y)_2Te_3$-$(Zn_{1-x}Cr_x)Te$, and did not yield an improvement in the temperature at which QAHE was observed in comparison to dilute magnetically doped 3D TIs[16].

To minimise the interface potential when inducing magnetic order *via* proximity, necessitates the use of ferromagnetic insulators such as $CrI_3$[17], $CrGeTe_3$[18] and single-layer $MnBi_2Te_4$[19] that possess similar chemical and structural compositions to 3D topological insulators. A single layer of the intrinsic magnetic topological insulator $MnBi_2Te_4$ is particularly promising[20]. Single-layer $MnBi_2Te_4$ possesses a band gap exceeding 780 meV[21], and has long-range ferromagnetic order (arising from the $Mn^{2+}$ ions that contribute a 5 $\mu_B$ magnetic moment, with the moments coupled ferromagnetically within each layer) and a Curie temperature of 12 K[20]. Importantly, it is structurally similar to the well-known 3D TI $Bi_2Te_3$, with the in-plane lattice constant of $MnBi_2Te_4$ (0.433 nm)[20] almost identical to $Bi_2Te_3$ (0.438 nm)[22]. Furthermore, single layer $MnBi_2Te_4$ and $Bi_2Te_3$ can be grown with molecular beam epitaxy (MBE)[21,22] allowing the possibility to engineer new ferromagnet / TI heterostructures. This offers simplicity and control far greater than using 2D exfoliation and heterostructure stacking of different van der Waals materials such as $Bi_2Te_3$ and $CrI_3$ [23].

The near identical crystal and atomic structures of $MnBi_2Te_4$ and $Bi_2Te_3$ means that instead of an abrupt interface potential there is a magnetic extension of the $Bi_2Te_3$ TSS into the $MnBi_2Te_4$ magnetic layer as depicted in Fig. 1a. This strong interaction of the TI surface states with the $Mn^{2+}$ ions induces a significant exchange splitting in the TSS of the topological insulator thin film and opens a large gap. This type of magnetic extension was first demonstrated on the top surface of a FMI/TI heterostructure comprising 15 quintuple-layer (QL) $Bi_2Se_3$ with a single layer of $MnBi_2Se_4$ on the top surface, which yielded a bandgap ~100 meV [24]. However, with only the top TSS gapped and a 15 QL TI layer, the bottom surface would remain gapless and unlikely to yield the QAH effect. A new theoretical proposal showed that if both the top and bottom TSS of an ultra-thin $Bi_2Te_3$ film are gapped by MBT layers and the inverted band structure is preserved, chiral spin-polarized edge states and a wide bandgap QAH insulator can be formed.[25] In this case, depending on the thickness of the $Bi_2Te_3$ in the heterostructure the bandgap is predicted to be between 38 meV for 1QL $Bi_2Te_3$ and 60 meV for 4 or 5 QL $Bi_2Te_3$[25].

In this work, we demonstrate the epitaxial growth of such a heterostructure which comprises 4 QL $Bi_2Te_3$ sandwiched between two single septuple-layers (SL) of $MnBi_2Te_4$ ($[MnBi_2Te_4]_{1SL}$-$[Bi_2Te_3]_{4QL}$-$[MnBi_2Te_4]_{1SL}$). For simplicity, for the remainder of the paper, we will refer to this structure as MBT/BT/MBT. We then utilise angle-resolved photoelectron spectroscopy (ARPES) at 8 K, which is below the predicted Curie temperature, $T_c$=12 K of 1 SL $MnBi_2Te_4$ to measure the electronic band structure and size of the induced exchange gap. We observed strong band anisotropy and large hexagonal warping of the massive Dirac Fermion, with a gap of 75 ± 15 meV, which is well above room temperature thermal broadening and appears to be a promising avenue towards measuring QAH effect at elevated temperatures.

## Results & Discussion

High-quality ultra-thin MnBi$_2$Te$_4$ and Bi$_2$Te$_3$ films and the MBT/BT/MBT heterostructure were grown using molecular beam epitaxy (MBE) on Si (111)- 7 × 7 substrates. These growths followed established growth procedures that allow for MnBi$_2$Te$_4$ and Bi$_2$Te$_3$ to be grown down to single layer thickness[21,22]. Further details can be found in the Methods section. Figure 1b shows the crystal structure of the MBT/BT/MBT heterostructure, with the magnetic moments from Mn atoms labelled with red arrows. The high quality of the films is confirmed from Reflection High Energy Electron Diffraction (RHEED) patterns shown in Figure S1. The sharp, bright streaks indicate that the film is of uniform thickness and high crystallinity. Angle-integrated core level photoelectron spectroscopy was taken on the MBT/BT/MBT heterostructure at $hv$=100 eV shown in Fig. 1c, and shows the expected Mn 3$p$, Bi 5$d$ and Te 4$d$ components. However, two distinct Te 4$d$ spin-split components separated by 0.7 eV are observed (represented with green and purple shading). The peak positions of the two Te 4$d$ components are consistent with two chemical binding environments, one from 1 SL MnBi$_2$Te$_4$ and another from the underlying 4QL Bi$_2$Te$_3$, as these binding energy positions are consistent with core level spectra taken independently on a 1 SL MnBi$_2$Te$_4$ film and a 4 QL Bi$_2$Te$_3$ film as shown in Fig. S2.

Figure 1d - f show ARPES spectra taken on 1SL MnBi$_2$Te$_4$, 4QL Bi$_2$Te$_3$, and the MBT/BT/MBT heterostructure respectively. Each spectrum is overlaid with the corresponding DFT calculated band structure. For 1SL MBT, only a M-shaped valence band is present below the Fermi level indicating a bandgap greater than 780 meV[21]. For 4QL BT, the $n$-type doping and band dispersion are consistent with previous reports[22] where the Dirac cone with Fermi velocity along ΓM, $v_F = 4.1 \pm 0.5 \times 10^5$ m/s is buried within the valence band and appears gapless within our experimental resolution, which is consistent with DFT that predicts 4 QL BT is gapless.[26] Additional high-resolution spectra, and Fermi surface map along with overlaid density functional theory (DFT) data can be found in Fig. S3. In the MBT/BT/MBT heterostructure, the band structure away from the Fermi energy (>1eV) is remarkably similar to that of 4QL BT (see Fig.S4 for the full valence band) while the bands near the Dirac region are noticeably different. As shown in the DFT calculated bands in Fig. 1f, unlike in BT, the Dirac cone on the surface of the heterostructure is elevated above the M-shaped valence band, and the Fermi velocity of the Dirac electron band away from the Dirac region along ΓM increases to $v_F =$

$5.0 \pm 0.5 \times 10^5$ m/s. This is significantly larger than that of 4QL BT ($4.1 \pm 0.5 \times 10^5$ m/s), and comparable to the Fermi velocity of 5 SL MnBi$_2$Te$_4$[21]. DFT calculated orbital projected band structures in Fig. S5 show that the bands in the MBT/BT/MBT heterostructure near the Dirac region are comprised of *p*-orbitals from Bi and Te atoms, with the Mn *d*-bands lying at much higher binding energy. This confirms that the major role of the Mn$^{2+}$ ions is to modify the crystal potential and produce an effective magnetic field in the out-of-plane direction, with negligible *p-d* hybridisation near the Fermi level.

We now turn our attention to the Dirac region of the MBT/BT/MBT heterostructure to investigate the magnetic coupling of the TI surface state, and whether a gap is opened. Figure 2a shows the raw ARPES spectra of the MBT/BT/MBT thin film taken at *hv*= 40 eV. Whilst there is some spectral weight near Γ in the Dirac point region that is due to Te-orbital-related matrix elements effects (more prominent between *hv*=47-60 eV, see Fig. S6 in SI), the system appears to be gapped. To determine the bandgap accurately, we analyse the energy distribution curves (EDC) around the Γ point. Figure 2b shows an EDC curve extracted at the Γ point in Fig. 2a. There are three distinct peak features, the first is located just below the Fermi level and corresponds to the bulk conduction band (BCB). The second and third peaks occur in the Dirac point regime at 0.21 eV and 0.28 eV. We attribute these two peaks to the Dirac electron band and hole band respectively, with the stronger peak at 0.28 eV attributed to the hole band top because of the higher density of states in the nearly flat band near Γ which we depict schematically in the top right corner in Fig. 2b. This clear peak splitting in the EDC spectra confirms there is a bandgap, and to confirm the size of the bandgap we fit the EDC profile with three individual Lorentzian line shapes. This fitting yields the bottom of the electron band to be 0.205 eV and the top of the hole band to be 0.280 eV, resulting in a band gap of 75 ± 15 meV. In Fig.2c, we present a stacking plot of the EDC curves taken within the k range of ±0.1 Å$^{-1}$ around Γ, with the EDC at Γ represented as a black curve. The blue points are peak positions of Dirac bands obtained by fitting these EDC curves with the same method used in Fig. 2b. Figure 2d plots the ARPES spectra, overlaid with the EDC analysis band dispersion from Fig. 2c (red points) and DFT band structure, showing excellent agreement and reproduces the key band dispersion features. Photon energy dependent ARPES spectra between 40-65 eV at 8 K were also analyzed using the EDC peak fitting method to confirm the 2D nature of the heterostructure and that the massive Dirac gap does not change with photon energy. Figure 2e plots bandgap as a function of photon energy, and across all photon energies the fluctuation of the band gap is within our energy resolution of ±15 meV, and the Fermi velocity equal to $5.0 \pm 0.5 \times 10^5 m/s$. Confirming the 2D nature of the band gap.

In Figure 3 and 4 we now turn to the anisotropy and band warping in the MBT/BT/MBT heterostructure by fitting to a gapped Dirac band model, which self-consistently reassures the band gap value determined from the EDC analysis described above. In Fig. 3a and b we plot the ARPES constant energy contours and the DFT calculated iso-energy contours respectively and highlight the constant energy contours at $E_F$ (Fig. 3c) and 0.4 eV (Fig. 3d). Overall, the agreement between experiment and theory is excellent. In both cases when approaching the massive Dirac gapped region from the Fermi level there is a clear evolution in the Fermi surface from hexagram to hexagon, that shrinks in the gapped region to a point which corresponds to the bottom of the electron band at ~0.2 eV, then at 0.25 eV in the hole band the surface texture evolves back into a hexagram upon moving further into the hole band. The overall warping away from the Massive Dirac region is reminiscent of Bi$_2$Te$_3$[3].

The band anisotropy in the Dirac electron band and the warping strength can be explained and modelled by adopting and modifying the Dirac Hamiltonian model of Bi$_2$Te$_3$ proposed by Fu[27]. We include an energy gap to describe the exchange coupling of the TSS with the exchange field $\Delta_e\sigma_z$, where $\sigma_z$ is the Pauli matrix from the top and bottom surface MBT SLs:

$$\epsilon(k) = D - \sqrt{(\Delta_e + \lambda k^3 \cos(3\theta))^2 + v_F^2 k^2} \qquad (1)$$

where $D$ denotes doping level, $\Delta_e$ represents the exchange gap, $\lambda$ is the warping strength, $v_F$ the asymptotic Fermi velocity at large momenta away from the gapped region, $k$ is the wave vector and $\theta$ the polar angle with respect to the ΓK direction.

To examine the hexagonal warping strength, we first fit the band dispersion, blue points in Fig. 2c, of the electron band along Γ-M where $\lambda=0$ to equation (1). From the extracted energy distribution curves (EDC) and momentum distribution curves (MDC) we perform a linear fit of the data away from the gapped region to obtain $v_F = 5.0 \pm 0.5 \times 10^5$ m/s, which also yields a value for the doping, $D$=0.28eV that corresponds to extrapolating the linear fit to k=0. Based on this $v_F$ and $D$ we then set $\Delta_e$ as a fitting parameter to yield $\Delta_e$=78 meV. This $\Delta_e$ value obtained from fitting to the gapped Dirac model is consistent with the gap value of 75 meV obtained from the EDC fitting method. By fixing $\Delta_e$ and $v_F$, we now fit the Γ-K direction to determine the hexagonal warping strength, $\lambda$=418 eV· Å$^3$, which is almost twice as strong as the strength in Bi$_2$Te$_3$(~250 eV· Å$^3$ [27]). We overlay the extracted data points from EDC peak fitting (circles) and the model (solid line) to the ARPES spectra in Fig. 4a and b for Γ-M and Γ-K respectively, showing excellent agreement. Furthermore, we overlay our model with the Fermi surface data in Fig. 4c and reproduce the key features observed experimentally.

We now proceed to confirm the magnetic origin of the band gap in the MBT/BT/MBT heterostructure by demonstrating the observation of band asymmetry as a consequence of broken TRS known as the exchange-Rashba effect[28]. The interplay between the hexagonal warping and magnetization in our MBT/BT/MBT heterostructure is also expected to result in a band asymmetry along K′-Γ- K directions. In this case, the hexagonal warping term also introduces a non-zero *k*-dependent out-of-plane spin texture with eigenvalue of |s$_z$| reaching maxima at K points and zero at M points in the Brillouin zone. Because the warping term is odd under inversion, similar to Bi$_2$Te$_3$, in MBT/BT/MBT |s$_z$| at minus K point (K′) is opposite to s$_z$ at positive K[27]. If TRS is preserved, the band dispersion remains the same. However, introducing magnetization perpendicular to the sample surface, the exchange field causes the energy shift in the surface state where the energy of states in the ΓK′ direction with spin s$_z$ parallel to the field direction is lowered while the energy of state along ΓK is increased. As a result, the band dispersion in the surface state along K′-Γ-K direction is asymmetric but remains symmetric along M′-Γ-M because of vanishing out-of-plane spin s$_z$. In comparison, the Dirac bands in Bi$_2$Te$_3$ thin films and bulk crystals are symmetric about Γ in K′-Γ-K directions[3,29].

We look for evidence of this in our ARPES spectrum taken at 8 K and photon energy of 63 eV along M′-Γ-M and K′-Γ-K directions shown in Fig.4a and b respectively. Without further analysis, a strong dichroism is already visible in both spectra. To probe this predicted asymmetry, we plot momentum distribution curves (MDC) taken at 0.1 eV below $E_F$ for both M′-Γ-M and K′-Γ-K directions in Fig. 4d and e. We then fit the MDCs with two Lorentzian line shapes, for M′-Γ-M we obtain -0.049±0.01 Å$^{-1}$ and +0.05±0.01 Å$^{-1}$ and a negligible asymmetry of 0.001 Å$^{-1}$ in Dirac electron band, consistent with the theoretical prediction that along ΓM the Dirac bands are symmetric. The fitting for K′-Γ-K on the other hand reveals the position of the left Dirac electron band to be -0.058±0.01 Å$^{-1}$ while the right branch position sits at 0.043±0.01 Å$^{-1}$, yielding an asymmetry of 0.015 Å$^{-1}$, which is greater than our angular resolution of 0.01 Å$^{-1}$. This splitting in k of 0.015 Å$^{-1}$ and band velocity of 3.3 eV· Å, allows us to calculate an exchange field of ~ 50meV, which further corroborates our band gap value obtained from the EDC analysis mentioned above. Furthermore, the fitting also demonstrates that the peak area ratio along K′-Γ-K is 2.3, compared with 1.7 along M′-Γ-M. The linear dichroism in Γ-K direction has contributions from both broken mirror symmetry from the polarized synchrotron light electric field and magnetization. In Γ-M direction, with out-of-plane spin $s_z$ being almost zero, the dichroism originates largely from the light induced symmetry breaking. Therefore, the increase of 0.6 in the peak area ratio in Γ-K can be attributed to the magnetisation i.e. the exchange field couples to different $s_z$ components in left and right branches of Dirac electron band along Γ-K direction. These states then enter the initial state in dipole matrix element and result in the dichroism. To further confirm this, we examine the DFT calculated surface atom projected band structure in Fig. 4f and g, which also predicts this dispersion asymmetry. Finally, we can rule out quantum confinement as the origin of the band gap opening because the band dispersion in K′-Γ-K would have been symmetric under the preserved TRS. Thus, the band asymmetry we observed confirms the magnetic origin of the band gap.

## Conclusion

In conclusion, we have grown a designer FMI/TI/FMI van der Waals heterostructure which is a robust QAH insulator with a bandgap of 75 ± 15 meV. We observe a highly anisotropic massive Dirac Fermion resulting from the exchange-Rashba effect, which confirms the magnetic origin of the band gap due to broken TRS. The realisation of robust band gap in a designer heterostructure provides a crucial step towards QAHE and axion insulator phases at elevated temperatures. It also establishes such designer heterostructures as promising platforms for studying novel topological phases and for high-temperature lossless transport applications. Finally, it is an ideal platform for realising recently predicted resonant photovoltaic effects [30] that require both strong hexagonal warping and broken Kramer degeneracy.

# Methods

Thin films of MnBi$_2$Te$_4$, Bi$_2$Te$_3$ and MBT/BT/MBT were grown using molecular beam epitaxy (MBE) on *n*-type doped silicon (111) substrates under ultra-high vacuum. The samples were immediately transferred after growth to an interconnected ARPES chamber at HERS Beamline 10.0.1, Advanced Light Source (ALS), Lawrence Berkeley National Laboratory.

**MnBi$_2$Te$_4$/ Bi$_2$Te$_3$/ MnBi$_2$Te$_4$ thin film growth on Si (111) substrate**

Effusion cells were used to evaporate elemental Bi (99.999%), Te (99.95%) and Mn (99.9%) for growth. The deposition rates were measured with a quartz crystal microbalance before growth to ensure the flux ratio of Te to Bi and Mn is greater than 10:1. The silicon (111) substrate was flash annealed to achieve the 7 × 7 surface reconstruction observed in Reflective High Energy Electron Diffraction (RHEED). The substrate temperature was then kept at 230 °C during the growth process, as this temperature was determined to yield the highest quality growth based off oscillations in the RHEED intensity indicating layer-by-layer growth. 1SL MnBi$_2$Te$_4$ was achieved by first growing 1 QL of Bi$_2$Te$_3$, then a bilayer of MnTe was deposited in order to spontaneously form MnBi$_2$Te$_4$. The 1SL MnBi$_2$Te$_4$ was subsequently annealed in Te flux for 5-10 mins to improve crystallinity. Following this, 4QL Bi$_2$Te$_3$ was grown, followed by 1SL MnBi$_2$Te$_4$. The sample was then annealed in Te flux for 10 mins at 230°C to improve crystallinity. RHEED taken along the [11$\bar{2}$] and [110] directions are shown in Fig.S1. The sharp high-intensity streaks in the pattern indicates large area single crystal epitaxial growth.

**Angle Resolved Photoelectron Spectroscopy (ARPES) measurements**

Our ARPES measurements were performed at Beamline 10.0.1 with linear polarised synchrotron light and data was taken using a Scienta R4000 analyser at temperatures between 8K and 15K. The overall energy resolution is 15 meV. The angular resolution is 0.2° or momentum resolution approximately 0.01 Å$^{-1}$ for the photoelectron kinetic energies measured.

**Density Functional Theory Calculations**

The first-principles calculations are based on the density functional theory (DFT), using the projector augmented wave (PAW) method [31] as implemented in the Vienna ab initio Simulation Package (VASP) [32,33]. The generalized gradient approximation (GGA) with Perdew-Burke-Ernzerhof (PBE) realization [34] were adopted for the exchange-correlation potential. The plane-wave cutoff energy was set to be 400 eV. To account for the correlation effects for Mn 3d-electrons, we used the GGA+U method introduced by Dudarev et al. [35] and the value of $U_{eff} = U - J$ was chosen as 5.34 eV which was adopted from the work of Otrokov el al [25]. The van der Waals (vdW) corrections have been taken into account by making use of the DFT-D3 approaches [36].

The structure models for $(Bi_2Te_3)_4$ and $(MnBi_2Te_4)/(Bi_2Te_3)_{n=3}/(MnBi_2Te_4)$ heterostructures were constructed by stacking the $Bi_2Te_3$ quintuple layers or $MnBi_2Te_4$ septuple layers in the order of ABC sequence. For the $(MnBi_2Te_4)/(Bi_2Te_3)_{n=3}/(MnBi_2Te_4)$ heterostructures, the inversion symmetry is preserved. A vacuum layer of 20 Å thickness is added to avoid artificial interactions between periodic images. The in-plane lattice parameters of these slab structures were fixed to the experimental value of $Bi_2Te_3$, which is 4.3835 Å [37]. The Brillouin zone was sampled by the k-grid with a spacing of $2\pi \times 0.02 Å^{-1}$ within a Γ-centered sampling scheme for structural optimization and electronic structure calculations. Spin-orbit coupling was considered when performing the structural optimization. The energy and force convergence criteria were set to be $10^{-6}$ eV and $10^{-2}$ eV/Å, respectively. The surface-resolved band structures were plotted using PyProcar [38]. From the DFT results, we constructed the maximally localized Wannier functions (MLWFs) for Mn-d, Bi-p and Te-p orbitals using WANNIER90 [39-41] [11-13] and the iso-energy spectrum was plotted by means of WannierTools [42].

**Acknowledgements**

M.T.E. was supported by ARC DECRA fellowship DE160101157. M.T.E., C.X.T., Q.L. and N. M., acknowledge funding support from ARC Centre for Future Low Energy Electronics Technologies (FLEET) CE170100039. M.T.E., Q. L., C.X.T. acknowledge travel funding provided by the International Synchrotron Access Program (ISAP) managed by the Australian Synchrotron, part of ANSTO, and funded by the Australian Government. This research used resources of the Advanced Light Source, which is a DOE Office of Science User Facility under contract no. DE-AC02-05CH11231. S. A. Y. acknowledges funding from Singapore Ministry of Education AcRF Tier 2 (Grant MOE2019-T2-1-001).

**Author contributions** M.T.E. devised the experiments. Q. L. analysed the data with assistance from M. T. T. Q. L., C. X. T., performed the MBE growth at Monash University. Q. L., C. X. T., performed the MBE growth and ARPES measurements at the ALS with the support from J.H. and S.-K.M. The DFT calculations were performed by W. W., and S. A. Y.Y. Q. L., and M.T.E. composed the manuscript. All authors read and contributed feedback to the manuscript. We acknowledge Michael Fuhrer for fruitful discussions.

**Competing interests** The authors declare no competing interests.

FIGURES

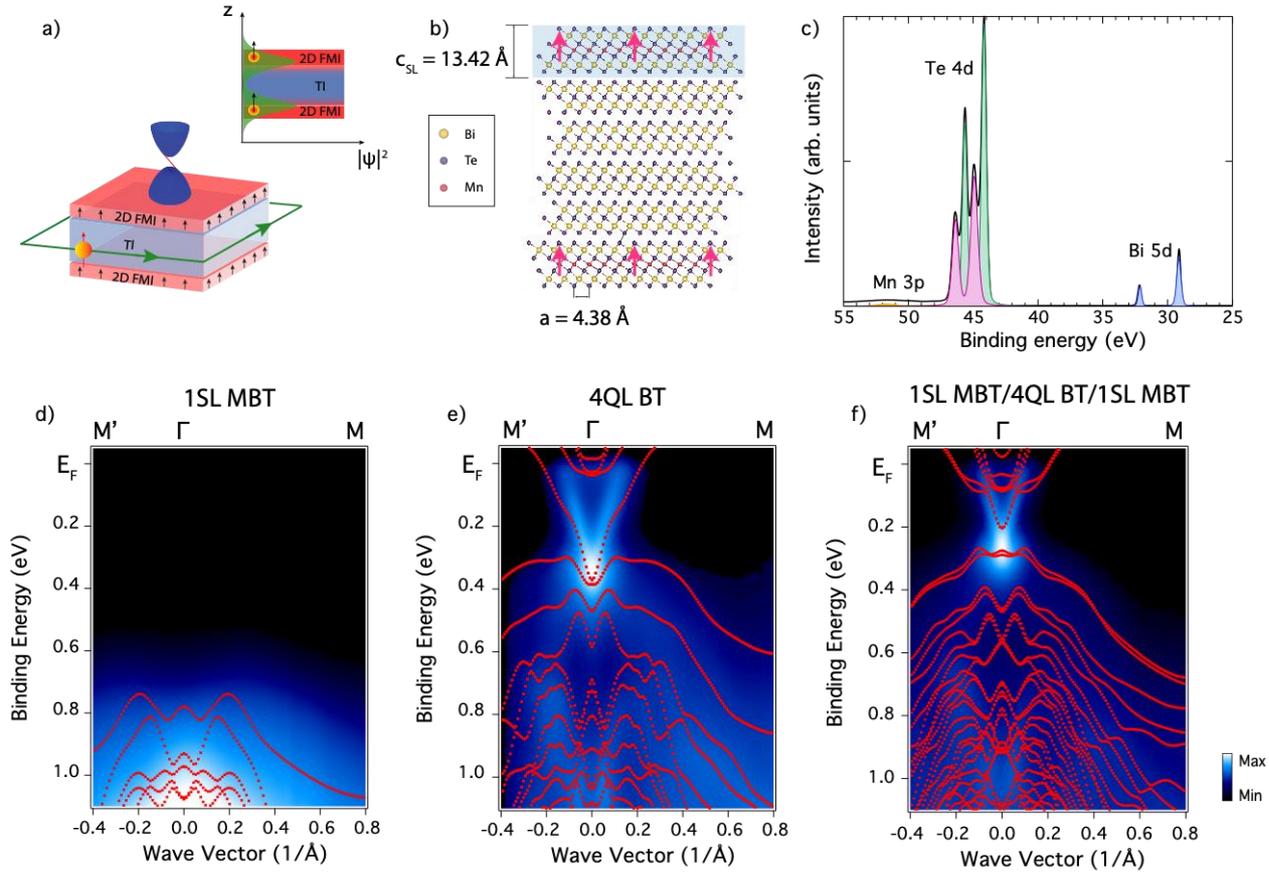

**Figure 1 A designer ferromagnetic insulator – topological insulator heterostructure, 1SL MBT / QL BT / 1SL MBT.** a) A schematic showing the realisation of the magnetic extension effect in the ferromagnetic insulator (FMI) – topological insulator (TI) heterostructure. Top right corner: wave functions of topological surface state (TSS) (green shaded curve) penetrate into the 2D FMI because of minimal lattice mismatch at the interface, resulting in a QAH insulator with chiral edge modes. b) Crystal structure of 4 quintuple layer (QL) $Bi_2Te_3$ (BT) sandwiched between two 1 septuple layers (SL) $MnBi_2Te_4$ (MBT) with the lattice constants labelled, magnetic moments on $Mn^{2+}$ marked by red arrows. The shadowed block indicates 1SL $MnBi_2Te_4$. c) Angle-integrated core level photoelectron spectroscopy measurement taken on the MBT/BT/MBT heterostructure at $h\nu$= 100 eV, showing the characteristic Mn *3p*, Bi *5d* and Te *4d* peaks. d)- f) ARPES spectra overlaid with DFT calculations along Γ-M direction for 1SL $MnBi_2Te_4$, 4QL $Bi_2Te_3$ and 1SL MBT/ 4QL BT/ 1SL MBT respectively.

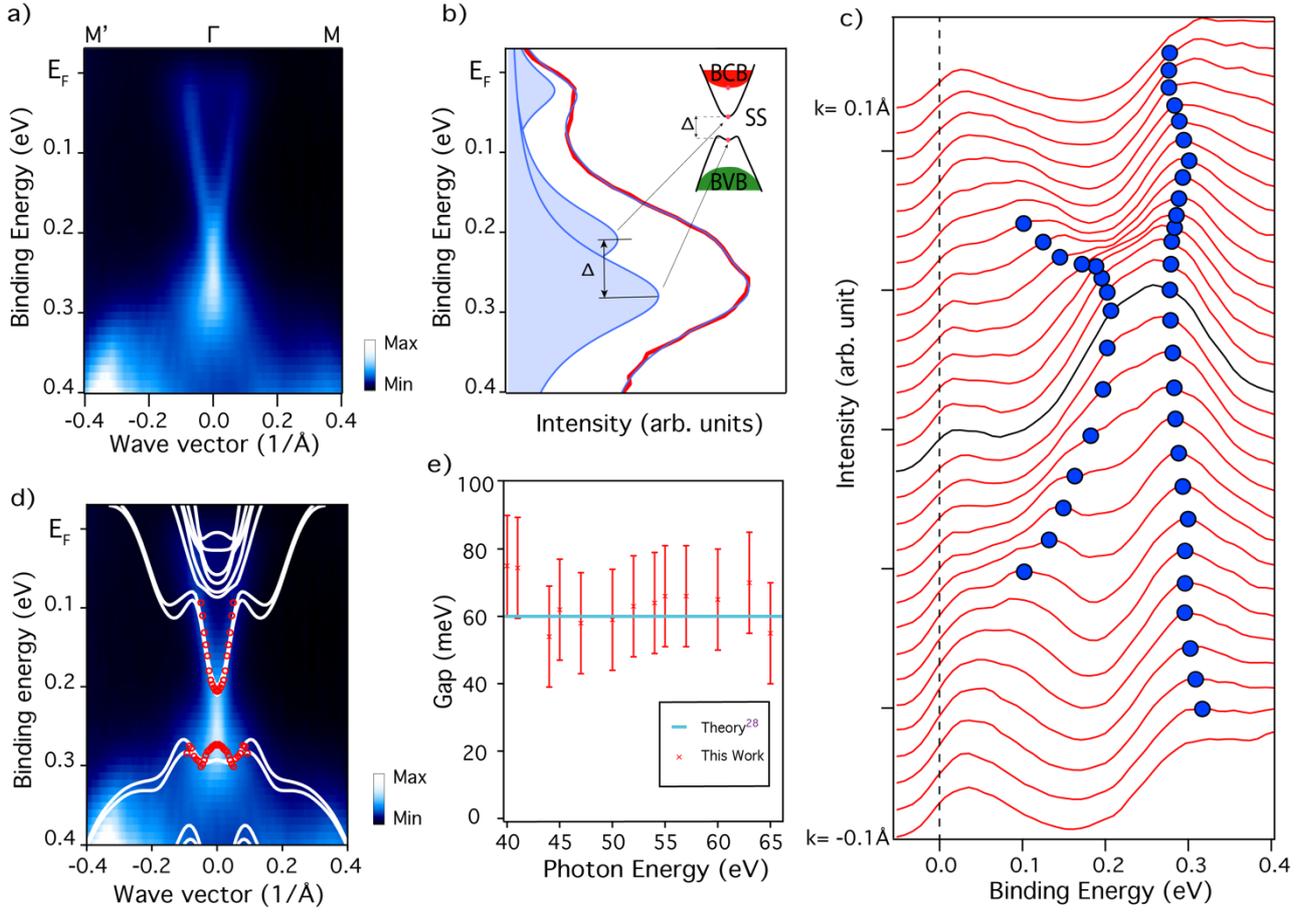

**Figure 2 Band gap analysis of MBT/BT/MBT heterostructure and its photon energy dependence**. a) ARPES spectra taken on MBT/BT/MBT at $h\nu=40$ eV at 8K with *p*-polarised light. b) EDC curve (red solid line) at the Γ point, with the fitted curve plotted as blue solid line. The shaded blue peaks represent individual Lorentzian peaks that correspond to the Dirac electron band, hole band and bulk conduction band (BCB). The separation of the hole band and Dirac electron band corresponds to a band gap, $\Delta=75 \pm 15$ meV. c) Stack plot of EDC curves taken within the wave vector range $\pm 0.1$ Å$^{-1}$. The EDC curve at Γ is highlighted in black and the Blue points are the peak positions obtained from EDC peak fitting analysis. The black dashed line marks the position of the Fermi edge. d) Same ARPES spectrum as in a) but overlaid with DFT calculated bands (white curves) and EDC analysis peaks (red open circles), showing excellent agreement. e) Band gap obtained from EDC analysis as a function of photon energy, the overall band gap value is consistent with the DFT theory plotted as a horizontal blue line.[25] An average band gap of $64 \pm 15$ meV has been obtained from e).

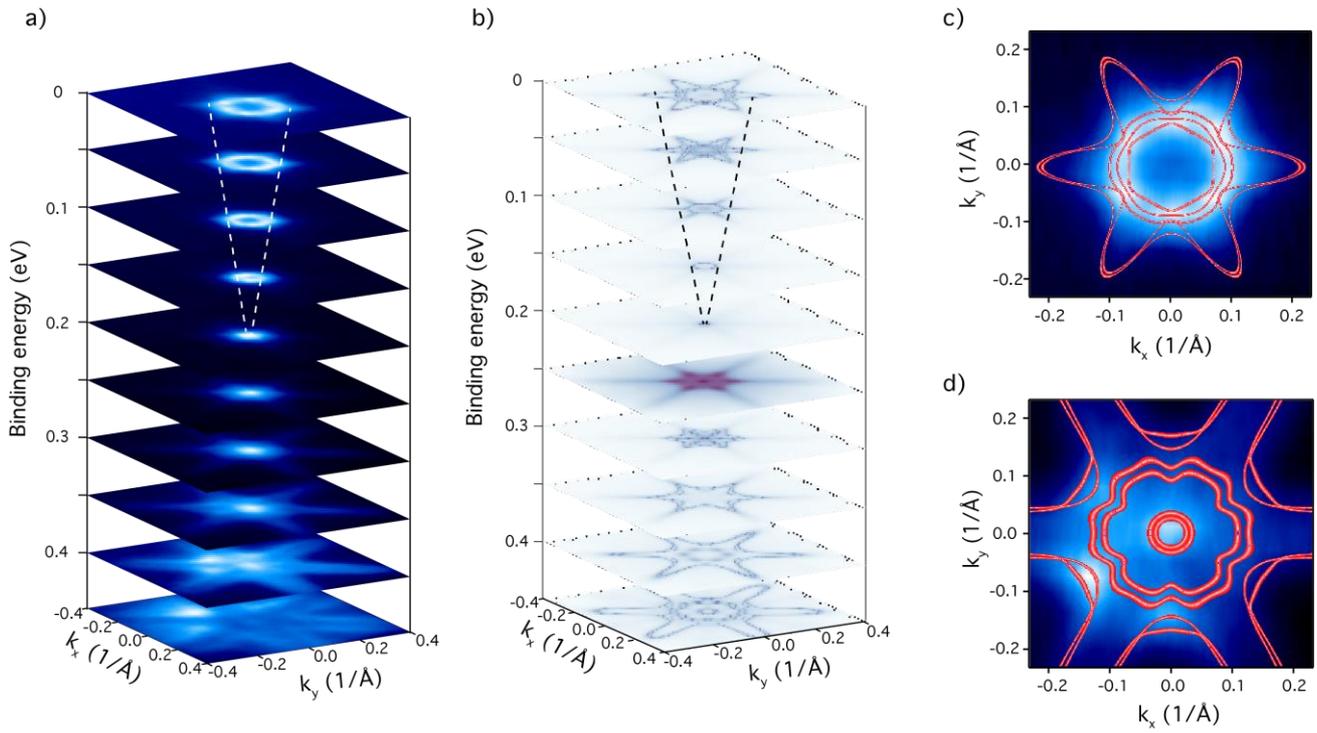

**Figure 3 Strong hexagonal warping in the MBT/BT/MBT heterostructure.** a) Iso-energy scans taken on MBT/BT/MBT across the gap region between the Fermi level and 0.4 eV. b) DFT calculated iso-energy scans in the same energy range showing excellent agreement. The dashed lines of the Dirac electron band are a guide to the eye. c) Fermi surface (FS) and d) iso-energy cut at 0.4 eV binding energy overlayed with DFT calculated contours in red. The angular intensity distribution shows three-fold symmetry which agrees with the FS calculated by DFT.

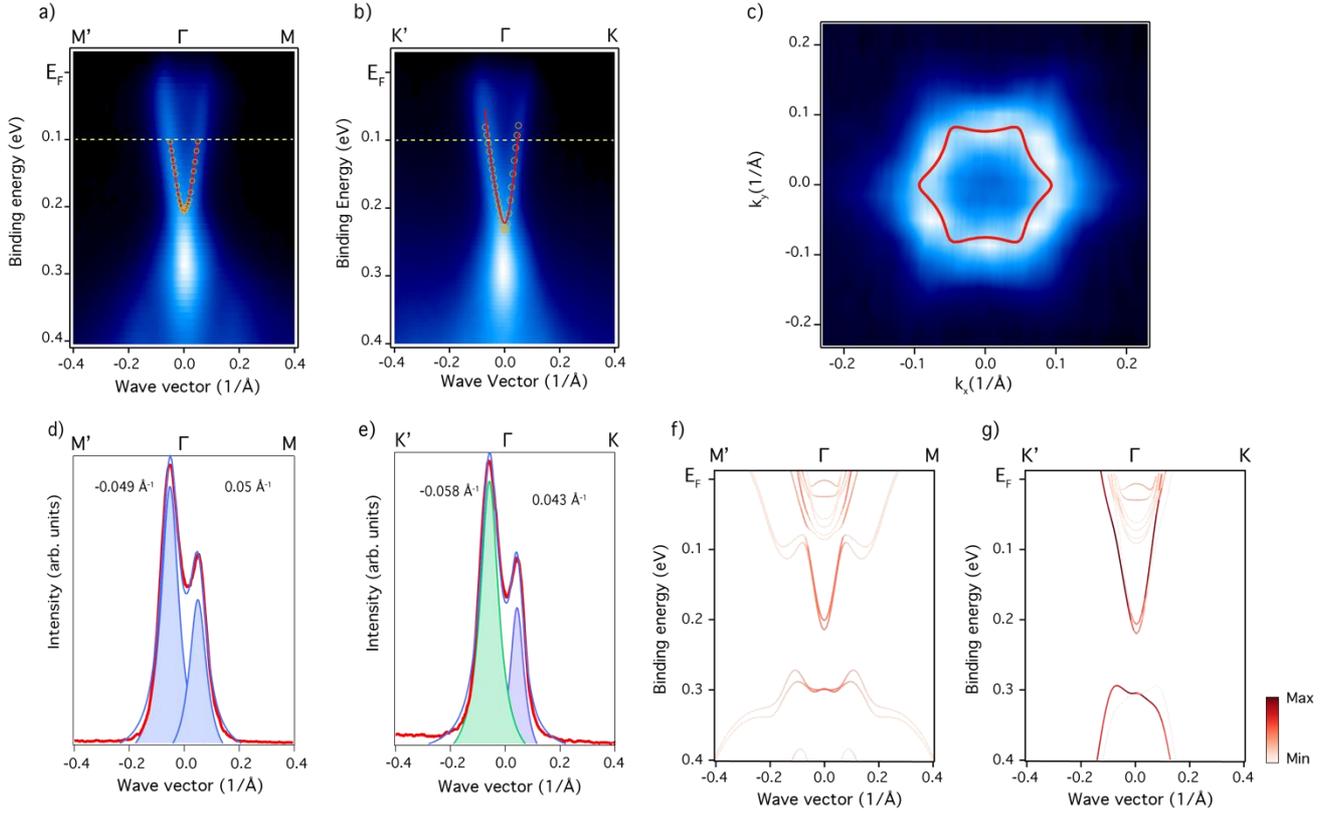

**Figure 4 Demonstration of time-reversal symmetry breaking and the exchange Rashba effect.** a) ARPES spectrum taken along $M'$-Γ-M direction at 63 eV photon energy, 8K with *p*-polarised light. The band is symmetric about Γ due to the preserved mirror symmetry and vanishing $s_z$ along ΓM direction. b) ARPES spectrum taken along $K'$-Γ-K at 63eV photon energy, 8K with *p*-polarised light. c) The Fermi surface scan overlayed with the contour from the model. d) and (e) represent momentum distribution curves (MDC) taken at 0.1eV binding energy from a) and b) (yellow dashed line) respectively and are fitted with two Lorentzian peaks for comparison. A clear asymmetry is observed for $K'$-Γ-K after careful calibration of the Γ point. f) and g) DFT calculated top surface atom projected band structure along $M'$-Γ-$M$ (symmetric) and $K'$-Γ-$K$ (asymmetric) respectively.

# Supplementary Information

# Large bandgap quantum Anomalous hall insulator in a designer ferromagnet-topological insulator-ferromagnet heterostructure

Qile Li[1,2,3], Chi Xuan Trang[1,2], Weikang Wu[4,5], Jinwoong Hwang[6], Nikhil Medhekar[2,3], Sung-Kwan Mo[6], Shengyuan A. Yang[4], Mark T Edmonds[1,2*]

[1]School of Physics and Astronomy, Monash University, Clayton, VIC, Australia.

[2]ARC Centre for Future Low Energy Electronics Technologies, Monash University, Clayton, VIC, Australia.

[3]Department of Materials Science and Engineering, Monash University, Clayton, VIC, Australia

[4]Research Laboratory for Quantum Materials, Singapore University of Technology and Design, Singapore 487372, Singapore

[5] Division of Physics and Applied Physics, School of Physical and Mathematical Sciences, Nanyang Technological University, Singapore 637371, Singapore

[6]Advanced Light Source, Lawrence Berkeley National Laboratory, Berkeley, CA, USA.

* Corresponding author: mark.edmonds@monash.edu

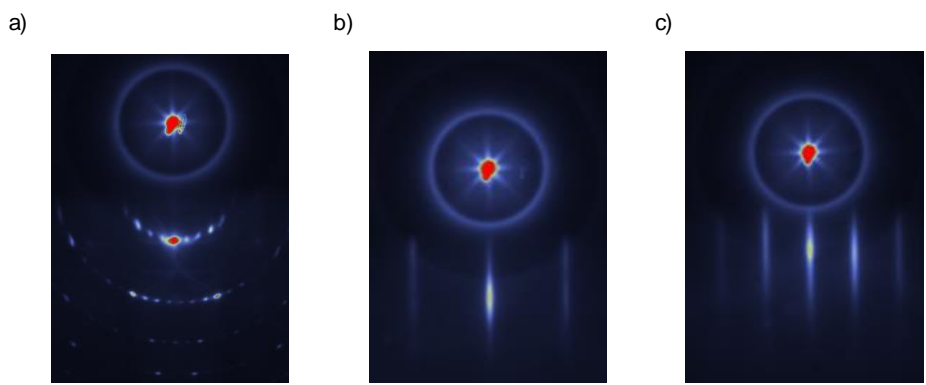

**Figure S1** Reflective High Energy Electron Diffraction (RHEED) pattern obtained at 15kV, 1.5A. a) Clean Si (111) substrate showing 7x7 surface reconstruction before growth. b) RHEED of MBT/BT/MBT heterostructure taken after growth along $[11\bar{2}]$ direction and c) taken along [110] direction. The sharp high intensity streaks in the patterns indicate large area of single-crystal epitaxial thin films.

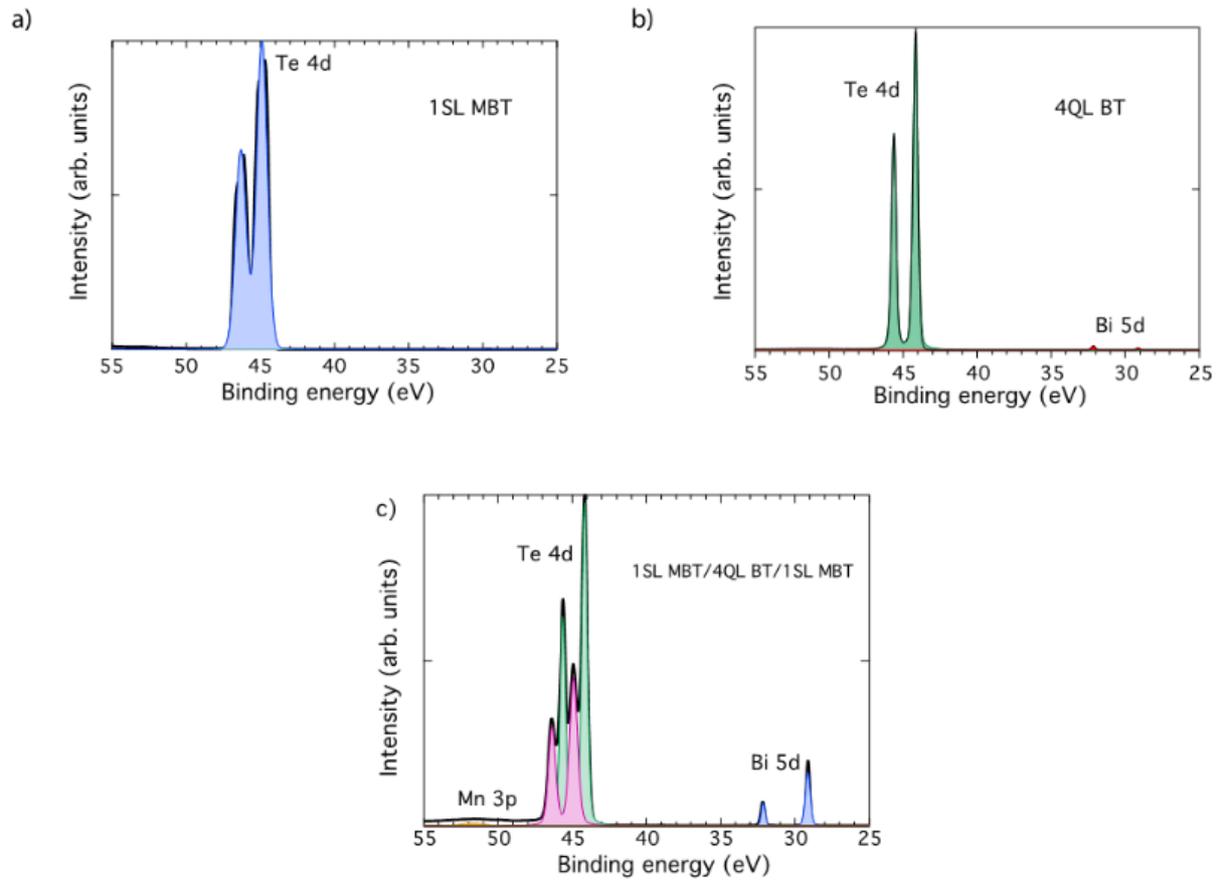

**Figure S2** X-ray Photoelectron Spectroscopy (Angle-integrated core level photoelectron spectroscopy) data taken at 100 eV photon energy to characterise the shallow Te 4d core levels of a) 1SL MBT on Si (111) substrate, b) 4QL BT on Si (111) substrate and c) MBT/BT/MBT heterostructure. The core level peaks are fit with Viogt line shapes with Shirley background. The positions of Te 4d doublet in a) 1SL MBT coincides with the 4d Te doublets in c) and positions of Te 4d peaks in 4QL BT coincides with the other doublet at lower binding energy. The presence of two Te 4d doublets can be attributed to Te in two bonding environments from MBT and underlying BT layer.

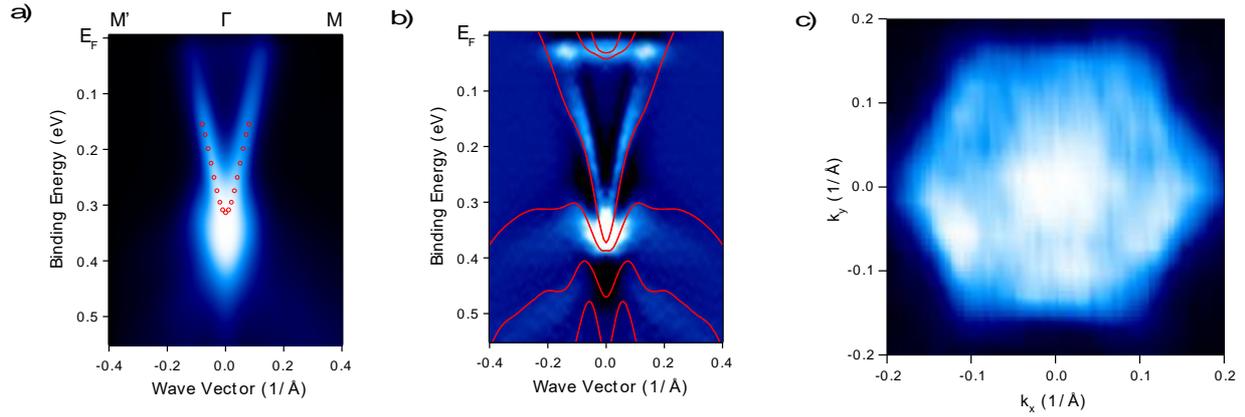

**Figure S3** High resolution scan of 4QL BT near the gap region. a) Original ARPES spectrum taken at 54eV photon energy with P-polarised light overlayed with Lorentzian peak positions as red points from EDC analysis. Fitting to these data points yields asymptotic velocity of 2.7 eV· Å. b) Second derivative enhanced spectrum overlayed with DFT band structure. c) Fermi surface of BT, showing different overall shape from the MBT/BT/MBT heterostructure with intensity from bulk conduction band in the centre but with similar hexagonal symmetry.

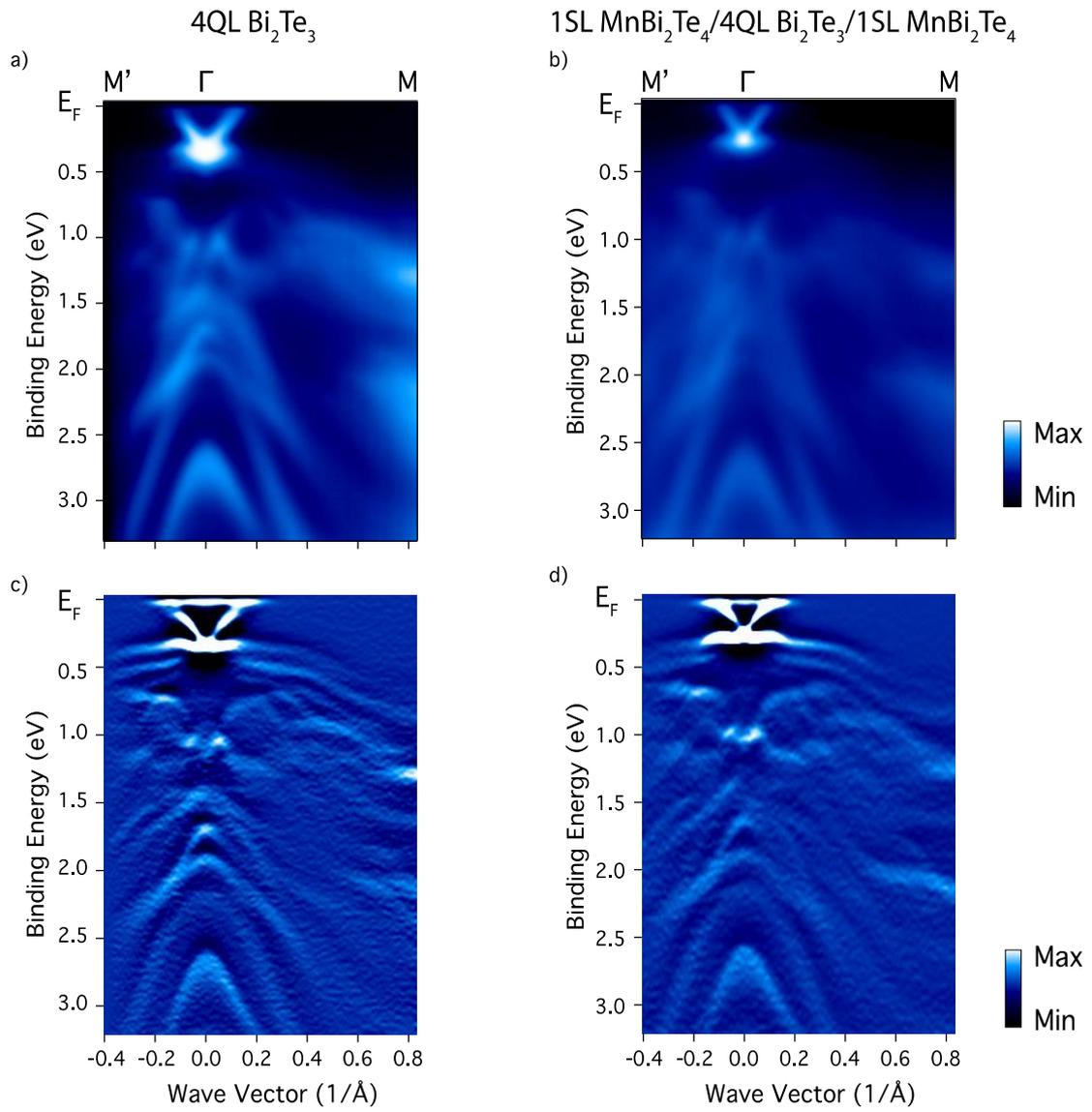

**Figure S4** Valence band scan of a) 4QL BT, b) 1SL MBT/ 4QL BT/ 1SL MBT and second derivative enhanced spectra c) and d) taken at 54eV photon energy under P-polarisation. The bands of the heterostructure at high binding energy resemble that of 4QL BT.

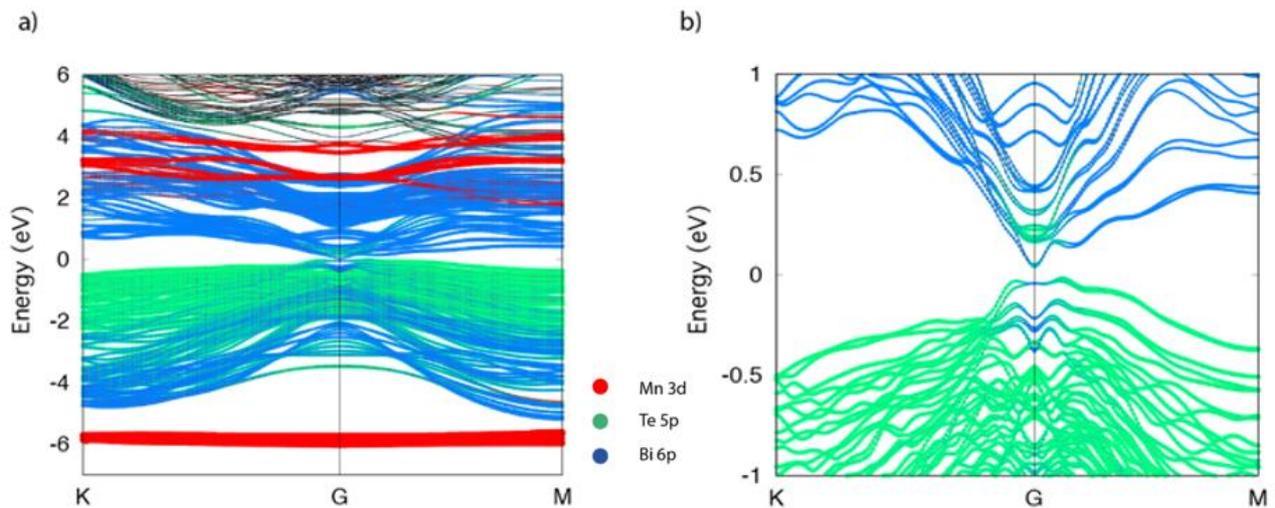

**Figure S5** Atom orbital decomposed band structure along K-Γ-M Directions. a) the overview of the band structure where Mn 3d orbitals are represented in red solid circles, Te 5p in green and Bi 6p in blue. b) band structure near the gap region taken from a). Near Γ point, the states consist of mixing of Bi 6p and Te 5p state. The Mn d bands are far below and above the band gap region and there is very little p-d hybridisation near the gap.

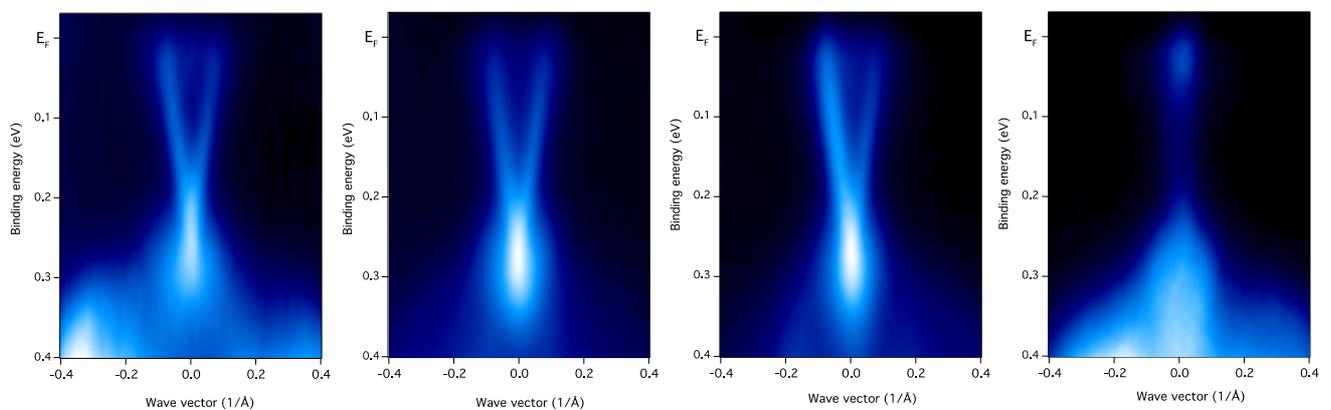

**Figure S6** Band structure showing Dirac bands near Fermi level taken at different photon energies: a) 40 eV, b) 50 eV, c) 60 eV, d) 70eV) with P-polarised light along ΓM direction. Dirac bands produce stronger signal at lower photon energy. At photon energy below 50 eV, the hole band top shows much weaker intensity.